\documentclass[12pt,preprint]{aastex}

\newcommand{\be}{\begin{equation}}
\newcommand{\ee}{\end{equation}}
\newcommand{\energy}{ {\cal E}} 
\def\lta{\,\raise 0.3 ex\hbox{$ < $}\kern -0.75 em
 \lower 0.7 ex\hbox{$\sim$}\,}
\def\gta{\,\raise 0.3 ex\hbox{$ > $}\kern -0.75 em
 \lower 0.7 ex\hbox{$\sim$}\,} 

\begin{document} 

\title{\bf EVOLUTION OF PLANETARY ORBITS WITH \\
STELLAR MASS LOSS AND TIDAL DISSIPATION} 

\author{Fred C. Adams$^{1,2}$ and Anthony M. Bloch$^{3}$}  

\affil{$^1$Physics Department, University of Michigan, Ann Arbor, MI 48109} 

\affil{$^2$Astronomy Department, University of Michigan, Ann Arbor, MI 48109} 

\affil{$^3$Math Department, University of Michigan, Ann Arbor, MI 48109} 

\begin{abstract}

Intermediate mass stars and stellar remnants often host planets, and
these dynamical systems evolve because of mass loss and tides.  This
paper considers the combined action of stellar mass loss and tidal
dissipation on planetary orbits in order to determine the conditions
required for planetary survival. Stellar mass loss is included using a
so-called Jeans model, described by a dimensionless mass loss rate
$\gamma$ and an index $\beta$. We use an analogous prescription to
model tidal effects, described here by a dimensionless dissipation
rate $\Gamma$ and two indices $(q,p)$. The initial conditions are
determined by the starting value of angular momentum parameter
$\eta_0$ (equivalently, the initial eccentricity) and the phase
$\theta$ of the orbit. Within the context of this model, we derive an
analytic formula for the critical dissipation rate $\Gamma$, which
marks the boundary between orbits that spiral outward due to stellar
mass loss and those that spiral inward due to tidal dissipation. This
analytic result $\Gamma=\Gamma(\gamma,\beta,q,p,\eta_0,\theta)$ is
essentially exact for initially circular orbits and holds to within an
accuracy of $\approx50\%$ over the entire multi-dimensional parameter
space, where the individual parameters vary by several orders of
magnitude.  For stars that experience mass loss, the stellar radius
often displays quasi-periodic variations, which produce corresponding
variations in tidal forcing; we generalize the calculation to include
such pulsations using a semi-analytic treatment that holds to the same
accuracy as the non-pulsating case. These results can be used in many
applications, e.g., to predict/constrain properties of planetary
systems orbiting white dwarfs.

\end{abstract}


\section{Introduction} 
\label{sec:intro} 

Planetary orbits around intermediate mass stars are subject to
evolution due to both stellar mass loss and tidal dissipation. In
systems where stars are actively losing mass, orbital evolution
represents a classic problem in solar system dynamics
(\citealt{jeans}; see also \citealt{hadji63,hadji66}). Similarly,
orbital changes due to tidal dissipation represent another classic
problem (e.g., \citealt{zahn,hut1981}). In both of these problems,
however, much of the previous work has focused on systems where both
bodies are stellar.  As outlined below, two classes of recent
observations motivate further studies of planetary systems that 
experience both stellar mass loss and tidal dissipation.  

One motivation for this work is the observed pollution of white dwarf
atmospheres \citep{gransicke2006,gransicke2007,melis}.  This
phenomenon occurs in approximately 25\% of the white dwarfs with
photospheric temperatures $T\lta25,000$ K, i.e., for those objects
where elements heavier than Helium are assumed to be recent external
pollutants (older contaminants would have sunk below the photosphere 
via gravitational settling). 
Within this collection of metal-enriched stars, a small
fraction ($\approx$10\%) have dust disks that are observable at infrared
wavelengths (\citealt{jura2007,farihi2009,farihi2010}; see also
\citealt{zuckerman2003,zuckerman2010}); a small fraction of the
systems with dust disks are observed to have gaseous metals in
addition to solids.  An emerging consensus interprets this
observational signature as arising from rocky planetesimals that are
tidally ripped apart, and subsequently accreted, by the white dwarf
star (starting with \citealt{jura}; see also \citealt{debes}). This
(inferred) presence of rocky planetesimals is often further
interpreted as evidence for planetary systems around these stellar
remnants.  Although no direct observations of planets have been made
in these systems, we nonetheless expect planets to orbit white dwarfs
(planets {\it have} been observed orbiting neutron stars;
\citealt{wolszczan}).  These host stars have undergone significant
mass loss between the main sequence and their current state as stellar
remnants, and sufficiently close planets would have experienced strong
tidal forces.  These systems thus pose the problem of planetary
survival in the face of stellar mass loss and tidal dissipation.

Related observations show that stars of intermediate mass often host
planets. Due to observational considerations, these planetary systems
are studied for host stars in their giant, post-main-sequence phases
of evolution (see \citealt{gettel} and reference therein). To
understand the currently observed orbital parameters of these planets,
and their expected future evolution, we again need to consider
planetary survival in systems with stellar mass loss and tidal
dissipation. Finally, this issue will eventually determine the
survival of our own planet \citep{schroder,spiegel}.

This work builds upon previous studies, which have considered
planetary survival in systems with substantial stellar mass loss
(\citealt{verasetal,verastout,akb2013}, hereafter AAB2013) and 
possible accretion of closer planets subjected to tidal forces
\citep{villaver07,villaver09,nordhaus2010,kunitomo,nordhaus2013,mustvillaver}.
Whereas the majority of this previous work was carried out
numerically, this contribution provides analytic results, with a focus
on the boundary in parameter space between systems where planets
survive and those where planets are accreted. Our main result is an
analytic determination of the this boundary (Section~\ref{sec:model}),
which is in good agreement with numerical results
(Section~\ref{sec:numerical}), and can be generalized to include
radial pulsations of the star (Section~\ref{sec:pulsate}).

\section{Orbits with Mass Loss and Tidal Dissipation} 
\label{sec:model} 

\subsection{Model Equations} 

This section presents model equations for single-planet planetary
systems that experience both stellar mass loss and tidal dissipation
of the orbit. Mass loss is assumed to take place isotropically, so
that orbital angular momentum is altered only by tidal dissipation.
The planetary mass is assumed to be small compared to the stellar
mass, so that the planet acts as a test particle.

In dimensionless form, the (reduced) equation of motion for the radial
position $\xi$ of the planet can be written (see AAB2013) 
\be{d^2\xi\over{dt^2}}={\eta\over\xi^3}-{m(t)\over\xi^2},\label{basicradial}\ee
where $t$ is time, $m$ is the dimensionless mass, 
\be{m(t)}\equiv{M(t)\over{M_0}},\ee 
and where $\eta$ determines the (dimensionless) specific 
angular momentum of the planetary orbit,
\be\eta\equiv{J^2}/(GM_0a).\label{angdef}\ee
Here, $a$ is the initial semimajor axis and $M_0$ is the initial
stellar mass.  The dimensionless radial coordinate is defined by
$\xi=r/a$ and the dimensionless time variable is given in units of
$\Omega^{-1}=(a^3/GM_0)^{1/2}$. Orbits starting with zero eccentricity
have $\eta=1$, whereas eccentric orbits have $\eta=1-e^2<1$ ($e$ is the
initial orbital eccentricity).

The equation of motion (\ref{basicradial}) is augmented by our
prescription for stellar mass loss. Following numerous previous authors, 
we use a so-called Jeans model \citep{jeans,hadji66,verasetal}, where
the stellar mass loss rate obeys the differential equation 
\be{dm\over{dt}}=-\gamma{m^{\alpha}},\label{mdotrate}\ee 
where the index $\alpha$ specifies the model and the constant
$\gamma$ determines the mass loss rate at the beginning of the epoch
($t=0,m=1$). After defining the timescale $\tau=M_{\ast0}/\dot{M}_{\ast0}$, 
the mass loss parameter $\gamma$ can be written 
\be\gamma={1\over\tau}\left({a_0^3\over{G}M_{\ast0}}\right)^{1/2}\approx
1.6\times10^{-6}\left({\tau\over0.1{\rm Myr}}\right)^{-1}\left({a_0\over1
{\rm AU}}\right)^{3/2}\left({M_{\ast0}\over1M_\odot}\right)^{-1/2}.\label{gammascale}\ee
We also need the tidal dissipation equation, which describes how tidal
interactions between the star and planet lead to loss of angular
momentum and orbital decay. Here we adopt the parametric form
\be{d\eta\over{dt}}=-\Gamma{m^{-p}}\xi^{-q}.\label{basictides}\ee
The power-law index $q\approx7$, typical in tidal forcing, whereas the
index $p$ accounts for increases in stellar radius during the mass
loss epoch.  In the context of orbital decay, many expressions for
tidal dissipation are found in the literature. The general form of
equation (\ref{basictides}) encapsulates the basic physics (see below)
and can model most of the expected behavior arising from tidal 
dissipation. 

\subsection{Tidal Dissipation} 
\label{sec:tides} 

The form (\ref{basictides}) for the tidal dissipation term is motivated 
by previous treatments (\citealt{zahn,phinney}), where the 
semimajor axis and eccentricity of the orbit decay according to 
\be{{\dot{a}}\over{a}}=-c_1F\qquad{\rm and}\qquad{{\dot{e}}\over{e}}=-c_2F,\ee
where the function $F$ takes the form 
\be{F}\equiv{1\over{t_{con}}}{M_{env}\over{M_\ast}}\left(1+{M_P\over{M_\ast}}\right)
{M_P\over{M_\ast}}\left({R_\ast\over{a}}\right)^8+{\cal O}(e^2),\ee 
where $t_{con}$ is the convective timescale of the star, $M_{env}$ is
the envelope mass, and all other symbols have their usual meanings.
Since the energy of a Keplerian orbit $\energy\propto-1/a$ (for constant
stellar mass), we can model dissipation through the ansatz
\be{{\dot\energy}\over\energy}=-{{\dot{a}}\over{a}}=c_1F,\label{edotadot}\ee
where we use dimensionless time. The orbital energy $\energy$ has 
the time derivative 
\be{\dot\energy}={{\dot\eta}\over2\xi^2}-{{\dot{m}}\over\xi},\label{dedt}\ee 
where the first term arises from tidal dissipation and the second
arises from stellar mass loss (AAB2013). Equations (\ref{edotadot})
and (\ref{dedt}) together imply ${\dot\eta}=2\xi^2\energy{c_1}F$. After
making the substitution $a\approx\xi$ and $\energy\approx-m/2\xi$,
correct to leading order in eccentricity, the dissipation equation
becomes 
\be{\dot\eta}=-c_1{m}\xi{F(\xi)}.\ee
The convection timescale \citep{mustvillaver} is given by 
\be{t_{con}}=\left({M_{env}R_{env}^2\over{3}L_\ast}\right)^{1/3}\left({GM_\ast\over{a^3}}\right)^{1/2}\qquad\ee
$$\qquad\approx0.80\left({M_{env}\over1M_\odot}\right)^{1/3} 
\left({R_{env}\over1{\rm AU}}\right)^{2/3}\left({L_\ast\over10^4L_\odot}\right)^{-1/3} 
\left({M_\ast\over1M_\odot}\right)^{1/2}\left({a\over1{\rm AU}}\right)^{-3/2},$$ 
so that (dimensionless) $t_{con}$ is of order unity. 
The dissipation term becomes 
\be{\dot\eta}=-c_1{m}\xi{1\over{t_{con}}}{M_{env}\over{M_\ast}} 
{M_P\over{M_\ast}}\left({R_\ast\over\xi}\right)^8=-{\Gamma}m^{-p}\xi^{-q},\ee 
where $M_P\ll{M_\ast}$; the second equality defines the constant $\Gamma$ and 
the indices $(p,q)$. The index $q=7$ in this treatment, but similar forms, 
with alternative $q$-values, can be derived. The index $p$ depends on
the time/mass evolution of the stellar radius $R_\ast$. Although the
starting stellar radius increases with stellar mass, individual stars
grow larger as they lose mass. As a result, $R_\ast\propto{u^\nu}$, where 
previous results \citep{hurley,mustvillaver} suggest that $0\le\nu<1$, 
and hence $0\le{p}<7$. The dissipation term is thus expected to
be more sensitive to $\xi$, with weaker dependence on $m$. 
The constant $\Gamma$ then becomes 
\be\Gamma\equiv{c_1\over{t_{con}}}\left[{M_{env}\over{M_\ast}}{M_P\over{M_\ast}} 
\left({R_\ast\over{a}}\right)^8\right]_0\approx{M_P\over{M_{\ast0}}} 
\left({R_{\ast0}\over a_0}\right)^8,\ee 
where all quantities are evaluated at the start of the mass loss
epoch. For typical parameters, 
\be\Gamma\approx10^{-3}\left({M_P\over{M_J}}\right) 
\left({M_{\ast0}\over1M_\odot}\right)^{-1}\left({R_{\ast0}\over1{\rm AU}}\right)^8 
\left({a_0\over1{\rm AU} }\right)^{-8},\label{biggamscale}\ee 
where $M_J$ is the mass of Jupiter. 

\subsection{Change of Variables} 
\label{sec:varchange} 

Because the time variable does not appear explicitly in the equations
of motion (\ref{basicradial},\ref{mdotrate},\ref{basictides}), we can
use stellar mass as the independent variable, i.e., as the measure of
time. Since the mass $m=m(t)$ is a strictly decreasing function of
time, we work instead in terms of the inverse
\be{u}\equiv{1\over{m}}.\ee 
This effective time variable $u$ starts at $u=1$ and increases
monotonically. In terms of $u$, the mass loss equation
(\ref{mdotrate}) becomes 
\be{\dot u}={\gamma}u^\beta,\label{udotrate}\ee 
where $\beta=2-\alpha$ and $\gamma$ is the same as before. 

With this change of variables, the equations of motion take the form  
\be{\gamma^2}u^{2\beta}\left[{d^2\xi\over{du^2}}+{\beta\over{u}}
{d\xi\over{du}}\right]={\eta\over\xi^3}-{1\over u \xi^2},\label{xiforce}\ee
and 
\be{\gamma}u^{\beta-p}{d\eta\over{du}}=-\Gamma\xi^{-q}.\label{xidiss}\ee 
The corresponding energy $\energy$ of the system becomes 
\be\energy={1\over2}\gamma^2{u^{2\beta}}\left({d\xi\over{du}}\right)^2 
+{\eta\over2\xi^2}-{1\over{u}\xi}.\label{energyfun}\ee 
By definition, the dimensionless energy has starting value
$\energy=-1/2$.  The (effective) time derivative of the energy 
reduces to the form 
\be{d\energy\over{du}}={1\over{u^2}\xi}+{1\over2\xi^2}{d\eta\over{du}}
={1\over{u^2\xi}}-{\Gamma\over2\gamma}{1\over u^{\beta-p}\xi^{2+q}}. 
\label{energydq}\ee 
In the absence of tidal dissipation ($\Gamma=0$), the energy is an
increasing function of time. The planet becomes unbound if the energy
becomes positive. For $\Gamma\ne0$, the orbit can lose energy and the 
planet can spiral inward. 

\subsection{Analytic Solution} 

Given the equations of motion (\ref{xiforce}) and (\ref{xidiss}), we
consider an expansion where $\gamma\ll1$. To leading order in
$\gamma$, the equations of motion become
\be\gamma u^{\beta-p}{d\eta\over du}=-\Gamma\xi^{-q} 
\qquad{\rm and}\qquad\xi=u\eta+{\cal O}(\gamma^2).\ee
After eliminating $\xi$, we integrate to find the solution $\eta(u)$, 
\be\eta^{q+1}=\eta_0^{q+1}-{(q+1)\over(q+\beta-p-1)} 
{\Gamma\over\gamma}\left[1-u^{-(q+\beta-p-1)}\right].\label{leadingeta}\ee
The corresponding solution for $\xi(u)$ has the form 
\be\xi^{q+1}=u^{q+1}\left[\eta_0^{q+1}-{(q+1)\over(q+\beta-p-1)} 
{\Gamma\over\gamma}\right]+{(q+1)\over(q+\beta-1)}{\Gamma\over\gamma}u^{2-\beta+p}. 
\label{leadingxi}\ee
Since the indices generally obey the ordering $q+1>2+p-\beta$, the
first term dominates at late times (large $u$).  The sign of the term
in square brackets determines whether the planet ultimately spirals
inwards or outwards. The critical value of the dissipation parameter,
marking the boundary between accretion and survival, is thus given by
\be{\Gamma\over\gamma}\approx\eta_0^{q+1}{(q+\beta-p-1)\over(q+1)}.\label{estimate}\ee

Note that there is a mismatch in initial conditions: For this leading
order solution to have the correct starting value of the angular
momentum parameter ($\eta=\eta_0$), the initial value of the radial
variable must be $\xi=\eta_0$. As a result, the orbit cannot start at
an arbitrary phase. (This result makes sense: The leading order
solution ignores $\xi$-derivatives, thereby producing a lower-order
differential equation, which allows fewer initial conditions.)

For completeness, the change in energy as $u\to\infty$ becomes 
\be\Delta\energy=\int_1^\infty{du}{d\energy\over{du}}=\int_1^\infty{du}\left({1\over{u^2\xi}}
-{\Gamma\over2\gamma}{1\over{u^{\beta}}\xi^{2+q}}\right)={1\over2\eta_0}.\ee
%i.e., the maximum $\Delta\energy$ depends only on $\eta_0$. 

%Furthermore, $\Delta\energy$ is somewhat larger than the critical
%value $\Delta\energy=1/2$ required for the orbit to become unbound in
%the limit $t\to\infty$ ($u\to\infty$).

\begin{figure} 
\figurenum{1} 
{\centerline{\epsscale{0.90} \plotone{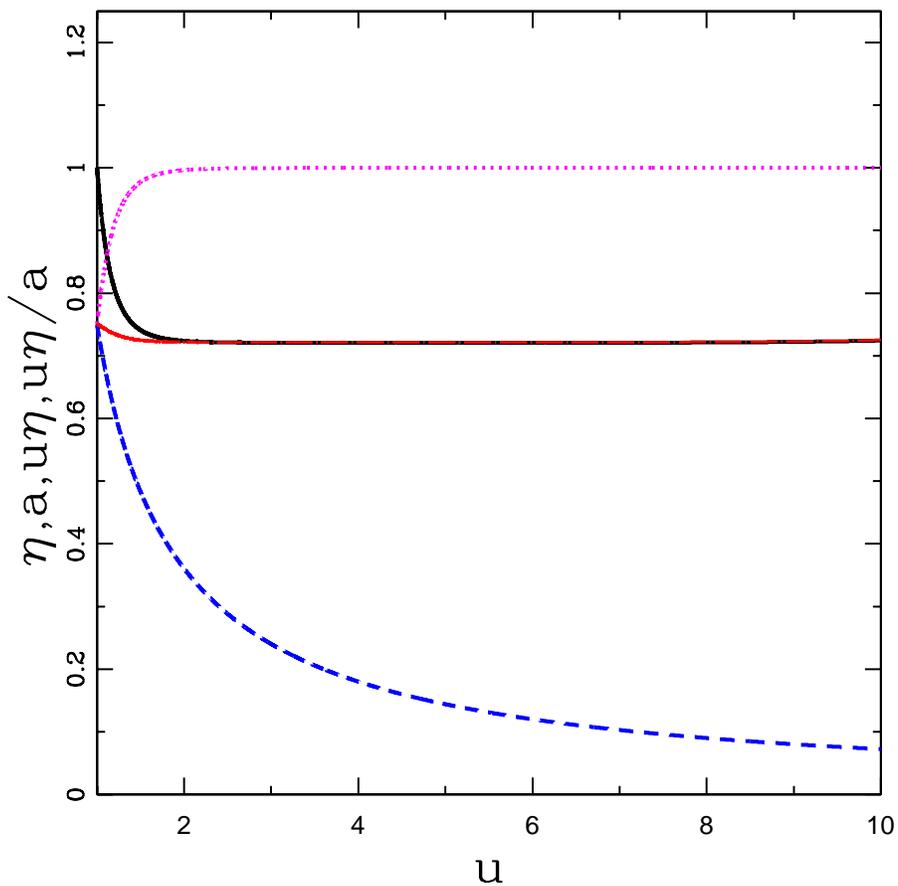} } } 
\figcaption{Field variables as a function of $u=1/m(t)$ for typical 
system with $\beta=2$, $q=7$, $\eta_0=0.75$, and $p=0$. The plot
shows the semimajor axis (black solid curve), angular momentum
parameter $\eta$ (blue dashed curve), the product $u\eta$ (red
dot-dashed curve), and the combination $u\eta/a$ (magenta dotted
curve). }
\label{fig:run} 
\end{figure} 

\begin{figure} 
\figurenum{2} 
{\centerline{\epsscale{0.90} \plotone{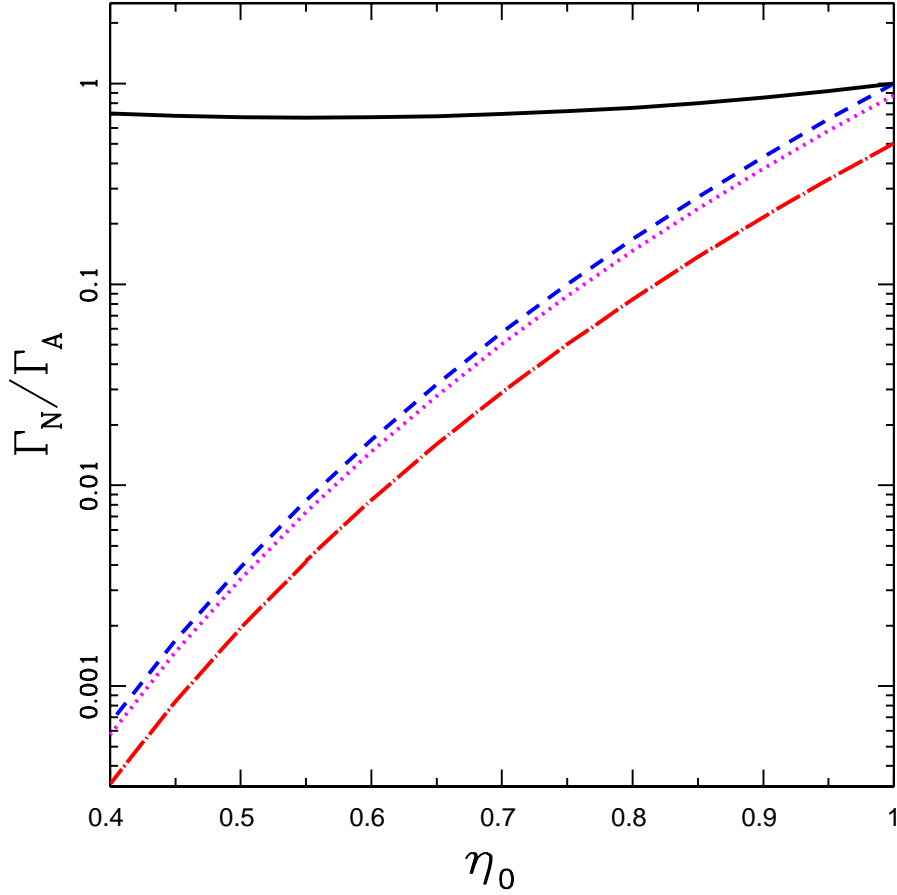} } } 
\figcaption{Critical value of tidal dissipation parameter $\Gamma$ 
that marks the boundary between planetary accretion and survival.
Results are shown for three cases with $q=7$ and varying indices
$(\beta,p)$. Solid black curve shows the ratio $\Gamma_N/\Gamma_A$
versus starting angular momentum parameter $\eta_0$ (same for all
models). For comparison, the analytic estimate itself
$\Gamma_A/\gamma=\eta_0^{q+1}(q+\beta-p-1)/(q+1)$ is plotted for
indices ($\beta=2,p=0$; blue dashed curve), ($\beta=1,p=0$; magenta
dotted curve), and ($\beta=2,p=4$; red dot-dashed curve). } 
\label{fig:gamveta} 
\end{figure}

\begin{figure} 
\figurenum{3} 
{\centerline{\epsscale{0.90} \plotone{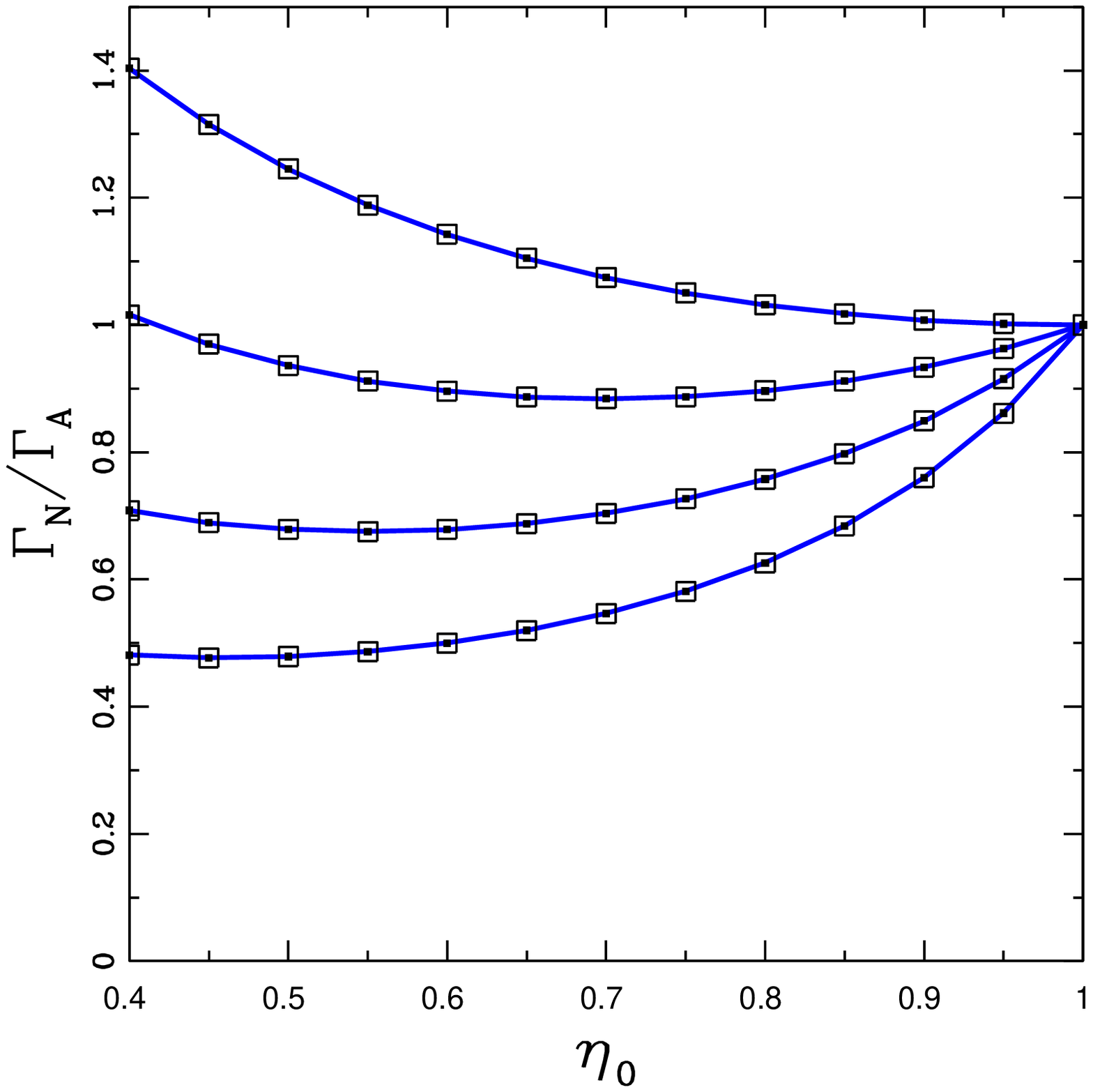} } } 
\figcaption{Critical value of tidal dissipation parameter 
($\Gamma_N/\Gamma_A$) for different indices $q=5-8$ (from top to   
bottom), plotted versus starting angular momentum parameter $\eta_0$.
All cases use $p=0$. Solid curves (blue) show results for $\beta=2$, 
whereas symbols show results for $\beta=1$ (smaller squares) and
$\beta=3$ (larger squares); the results are essentially independent 
of $\beta$. } 
\label{fig:qcompare} 
\end{figure} 

\section{Numerical Results} 
\label{sec:numerical} 

Equation (\ref{estimate}) provides an analytic estimate for the value
of $\Gamma$ the marks the boundary between systems where planets are
accreted and those where they survive. We denote this analytic value
as $\Gamma_A$.  In this section we numerically integrate the equations
of motion to determine the value $\Gamma_N$ that marks this boundary.
As shown below, the analytic estimates are quite good, in that the
ratio $\Gamma_N/\Gamma_A$ is always of order unity.

Individual orbits are integrated using a Bulirsch-Stoer alogrithm,
where the relative error per time step is $10^{-12}$; in the iteration
loop for the critical value of $\Gamma$, the relative error due to
lack of convergence is $\approx3\times10^{-5}$. Numerical errors are 
thus negligible. 

Figure \ref{fig:run} shows how the system variables evolve with time
for a typical model ($\beta=2,q=7,p=0,\eta_0=0.75$). The angular
momentum parameter $\eta$ steadily decreases with time, whereas the
semimajor axis $a$ and the product $u\eta$ approach constant values.
The combination $u\eta/a$ approaches unity. This type of evolution
occurs over a wide range of the parameter space (not shown).

Figures \ref{fig:gamveta} and \ref{fig:qcompare} show the ratio
$\Gamma_N/\Gamma_A$ as a function of the starting value of the angular
momentum parameter $\eta_0$. These calculations use $\gamma=10^{-4}$;
the results are indpendent of $\gamma$ provided that $\gamma\ll1$.
Figure \ref{fig:gamveta} also shows the right-hand-side of equation
(\ref{estimate}). Note that for $\eta_0<1$, this quantity decreases
rapidly, varying by several orders of magnitude for the cases shown
here. Although the $\Gamma_N$ value that marks the boundary varies by
several orders of magnitude, the ratio $\Gamma_N/\Gamma_A$ always
remains of order unity.

Our numerical exploration reveals the following general results: 
[1] For circular orbits with $\eta_0=1$, the ratio
$\Gamma_N/\Gamma_A=1$, so that the analytic estimate from equation
(\ref{estimate}) is exact. 
[2] The numerically-determined results for $\Gamma_N/\Gamma_A$ are
independent of the mass loss index $\beta$ (where we focus on the
range $\beta=1-3$).  Similarly, the results for $\Gamma_N/\Gamma_A$ 
are independent of the index $p$ that determines how tidal dissipation 
depends on changing stellar radius during the mass loss epoch. 
[3] All of the results shown here are carried out for orbits that
start at periastron and for small $\gamma=10^{-4}$.  Our numerical
work shows that the results are independent of the starting orbital
phase $\theta$ for sufficiently small $\gamma$, but order-unity
differences arise for larger $\gamma\gta0.001$. However, stellar mass
loss takes place over $0.1-1$ Myr, and planets near the border have 
$a\approx1$ AU, so the physically-relevant values lie in the range
$\gamma=10^{-6}-10^{-4}$. 
[4] The ratio $\Gamma_N/\Gamma_A$ depends weakly on the index $q$ that
determines how tidal dissipation depends on radial distance.  Figure
\ref{fig:qcompare} illustrates this behavior by plotting the ratio
$\Gamma_N/\Gamma_A$ versus the starting angular momentum parameter
$\eta_0$ for $q=5-8$. Although the critical value of $\Gamma$ itself
varies by several orders of magnitude over the parameter space
depicted by the figure, the ratio $\Gamma_N/\Gamma_A$ remains of order
unity for all $q$ (and is independent of $\beta$ and $p$).

\section{Effects of Radial Pulsation} 
\label{sec:pulsate}  

This section revisits the leading order solution with the inclusion of
non-monotonic variations in the stellar radius.  The radial pulsations
of the star often appear to be periodic, or nearly periodic
\citep{vasswood}.  Since the tidal dissipation term is proportional to
the stellar radius to a high power, even small changes can make a
difference.  To incorporate the radial variations, we multiply the
tidal dissipation term by a factor of the form 
\be{F}=1+bQ(t),\ee 
where $Q(t)$ is a quasi-periodic function with vanishing mean and
where the parameter $b$ is determined by the pulsation amplitude.
To leading order, the equations of motion have the form 
\be{\gamma}u^{\beta-p}{d\eta\over{du}}=-\Gamma\xi^{-q}\left[1+bQ(u)\right] 
\qquad{\rm and}\qquad\xi=u\eta+{\cal O}(\gamma^2),\ee
where the time dependence of the quasi-periodic function $Q$ is
written in terms of $u$. After eliminating $\xi$, as before, we 
integrate to find the solution $\eta(u)$, 
\be\eta^{q+1}=\eta_0^{q+1}-{(q+1)\over(q+\beta-p-1)} 
{\Gamma\over\gamma}\left[1-u^{-(q+\beta-p-1)}\right]  
-{b(q+1)\Gamma\over\gamma}\int_1^u{u^{-(q+\beta-p)}}Q(u)du.\label{etapulse}\ee
It is convenient to define the integral quantity 
\be{I}(u)\equiv\int_1^u{u^{-(q+\beta-p)}}Q(u)du.\ee 
Converting back to the function $\xi$, we have the solution 
\be\xi^{q+1}=u^{q+1}\left[\eta_0^{q+1}-{(q+1)\over(q+\beta-p-1)} 
{\Gamma\over\gamma}\right]+{(q+1)\over(q+\beta-p-1)}{\Gamma\over\gamma}u^{2-\beta+p} 
-{b(q+1)\Gamma\over\gamma}u^{q+1}I(u).\label{xipulse}\ee 

The indices generally obey the ordering $q+1>2-\beta+p$, and we
expect the integral $I$ to converge. At late times, corresponding to
large $u$, we want the solution $\eta(u)$ to approach zero
and the solution $\xi(u)$ to remain finite. These conditions imply
that the critical value of the tidal dissipation parameter is given by 
\be{\Gamma\over\gamma}=\eta_0^{q+1}{(q+\beta-p-1)\over(q+1)}
\left[1+b(q+\beta-p-1)I_\infty\right]^{-1},\label{condition}\ee
where we take the limit $u\to\infty$ to evalute the integral. 
The quantity $I_\infty$ is usually, but not always, positive. As a
result, the effect of radial pulsations is (usually) to reduce the
size of the tidal dissipation parameter needed to make planets spiral
inward and be accreted by the star. In other words, pulsations generally 
result in more tidal dissipation, compared to the average 
(recall that $Q(t)$ has zero mean by definition).

To illustrate the effects of pulsations, we consider the tidal
dissipation strength to vary with time as a sine function, i.e.,
\be{Q}(t)=\sin\omega{t},\label{qsine}\ee where the pulsation period
$P=2\pi/\omega$. For exponential mass loss ($\beta=1$), $p=0$, and
arbitrary $q$, we must evaluate the integral
\be{I_\infty}=\int_1^\infty{u^{-(q+\beta)}}Q(u)du=
\int_1^\infty{du}u^{-(q+1)}\sin\left[{\omega\over\gamma}\ln{u}\right],\ee
which reduces to the form
\be{I_\infty}={\gamma\omega\over\omega^2+q^2\gamma^2}.\label{intsine}\ee
As another example, consider the $Q(t)$ profile from equation
(\ref{qsine}) with constant mass loss rate ($\beta=2$) and index
$q=8$; the integral $I_\infty$ then becomes
\be{I_\infty}={40320(1-\cos[\omega/\gamma])-20160(\omega/\gamma)^2+1680(\omega/\gamma)^4
  -56(\omega/\gamma)^6+(\omega/\gamma)^8\over(\omega/\gamma)^9}.\label{intsine2}\ee
Equations (\ref{intsine},\ref{intsine2}) represent two typical cases;
one can evaluate $I_\infty$ for any pulse profile and choice of
indices (although the integral sometimes must be done numerically, 
and we require $q+\beta>p+1$ for convergence).

These results depend on $\omega/\gamma$, which typically falls in the
range $\omega/\gamma\approx50-150$ for post-main-sequence stars during
their mass-loss epoch \citep{vasswood}. In the limit where the pulse
frequency is large compared to the mass loss rate, $\omega\gg\gamma$,
the integral $I_\infty\to0$, as expected: In this limit, oscillations
of the stellar radius average to zero. For pulsations to produce an
appreciable effect, the system must experience non-trivial evolution
(mass loss) during the course of a pulse (to break the symmetry).
Notice also that $I_\infty\to0$ as $\omega\to0$, the limit where
pulsations effectively vanish.

\begin{figure} 
\figurenum{4} 
{\centerline{\epsscale{0.90} \plotone{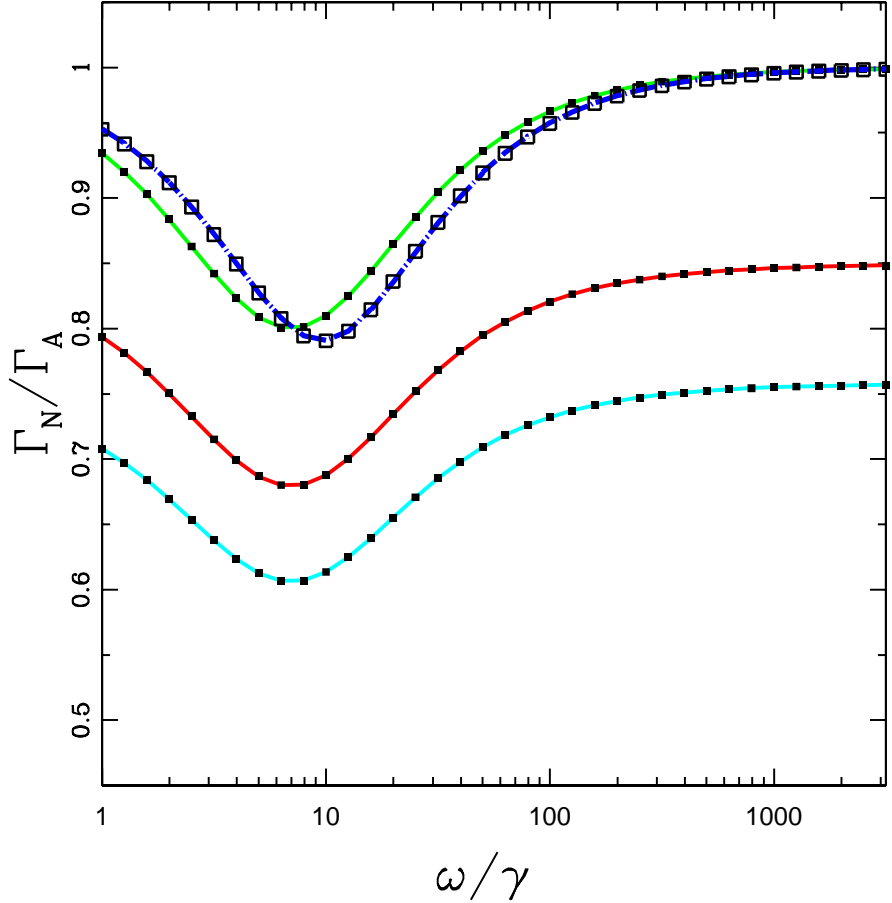} } } 
\figcaption{Critical tidal dissipation parameter ($\Gamma_N/\Gamma_A$), 
the boundary between planetary accretion and survival, plotted versus
pulsation frequency $\omega/\gamma$, for systems undergoing sinusoidal
pulsations (with amplitude $b=0.5$).  Solid curves show
numerically-determined results (for $\beta=1,q=7,p=0$), for initial
angular momentum $\eta_0=1$ (top green), 0.9 (middle red), and 0.8
(bottom cyan); solid squares show the corresponding analytic 
predictions. Dot-dashed (blue) curve shows the critical $\Gamma$ 
for constant mass loss rate $\beta=2$ (and $q=8$). } 
\label{fig:pulseone} 
\end{figure}

With the integral $I_\infty$ specified, the critical tidal damping
parameter is given by equation (\ref{condition}).  For initially
circular orbits, this result is exact: Figure \ref{fig:pulseone} shows
the numerically-determined value of the critical tidal damping
parameter for stars experiencing pulsations specified by equation
(\ref{qsine}), plotted versus frequency $\omega/\gamma$. The upper
curves correspond to initially circular orbits ($\eta_0=1$), with
$\beta=1$ (solid-green) and $\beta=2$ (dot-dashed-blue). The square
symbols show the analytically predicted results from equation
(\ref{condition}), using equations (\ref{intsine}) and
(\ref{intsine2}), respectively.  The lower curves show results for
$\eta_0=0.9$ (red) and 0.8 (cyan). Here, the square symbols represent
the numerical result with no pulsations (see Figure
\ref{fig:qcompare}) scaled using the pulsation correction factor
$[1+b(q+\beta-p-1)I_\infty]$.

\section{Conclusion}
\label{sec:conclude}  

Motivated by observations indicating that planets often orbit both
intermediate mass stars and stellar remnants, this paper considers the
combined effects of stellar mass loss and tidal dissipation on
planetary orbits. Mass loss allows planetary orbits to spiral outward,
whereas tidal dissipation acts to move planets inward. This work
determines the critical value of the tidal dissipation parameter that
marks the boundary between these two fates. To address this question,
we have developed a parametric model for orbital evolution that
accounts for stellar mass loss (equations [\ref{mdotrate},
  \ref{udotrate}]) and dissipation (equation [\ref{basictides}]).
These effects are specified by dimensionless rates ($\gamma,\Gamma$),
and indices that determine their functional forms ($\beta,q,p$). The
initial conditions are specified by the starting angular momentum
$\eta_0$ and orbital phase $\theta$ (although the results are 
independent of $\theta$ for $\gamma\ll1$). 

The main result of this work is an analytic expression for the
critical tidal dissipation rate $\Gamma$ that determines the boundary
between planetary survival and accretion (equation [\ref{estimate}]).
This formula specifies the dependence of the critical $\Gamma$ on
starting angular momentum $\eta_0$, mass loss rate $\gamma$, mass loss
index $\beta$, dissipation radial index $q$, and dissipation mass
index $p$ (see equation [\ref{estimate}]). The resulting expression is
accurate to $\approx50\%$ over entire 7-dimensional parameter space
$(\Gamma,\gamma,\beta,q,p,\eta_0,\theta)$, even though individual 
parameters vary over several orders of magnitude (see Figures 
\ref{fig:gamveta} and \ref{fig:qcompare}); moreover, the analytic
result is effectively exact for initially circular orbits. These 
results show that the most important variable in the problem is
the ratio $\Gamma/\gamma$, which can be written in scaled form, 
\be{\Gamma\over\gamma}\approx\left({\tau\over0.1{\rm Myr}}\right)\left({M_P\over{M_J}}\right) 
\left({M_{\ast0}\over1M_\odot}\right)^{-1/2}\left({R_{\ast0}\over1{\rm AU}}\right)^8 
\left({a_0\over2{\rm AU} }\right)^{-19/2},\label{ratscale}\ee 
where the timescale $\tau=M_{\ast0}/\dot{M}_{\ast0}$. Combining this
expression with the critical tidal dissipation parameter (equation
[\ref{estimate}]), we find the innermost orbits where planets survive,
\be\left({a_0\over2{\rm AU}}\right)\approx\eta_0^{2(q+1)/19}\left[{q+1\over{q+\beta-p-1}}
\left({\tau\over0.1{\rm Myr}}\right)\left({M_P\over{M_J}}\right)\right]^{2/19} 
\left({M_{\ast0}\over1M_\odot}\right)^{-1/19}\left({R_{\ast0}\over1{\rm AU}}\right)^{16/19}.\ee 

This work includes an important additional complication: evolved stars
often display pulsating variations in stellar radius, which produce
corresponding variations in tidal dissipation. Section
\ref{sec:pulsate} includes stellar pulsations in the formulation and
estimates the generalized critical value of the tidal dissipation
parameter (equation [\ref{condition}]); this semi-analytic expression
holds to the same accuracy as the non-pulsating case (Figure 
\ref{fig:pulseone}). 

Previous studies of planetary orbits with stellar mass loss indicate
that the product of mass and semimajor axis often remains nearly
constant, and this approximation ($am=constant$) is often used 
(\citealt{verasetal,mustvillaver}).  For purposes of finding the
critical tidal dissipation strength, however, this approximation
breaks down: The orbits of interest arise near the boundaries of
parameter space, where the planets are close to spiraling inward. But
if a planet spirals inward as the star loses mass, the product $am$
would decrease ($am\ne{constant}$).  A more applicable approximation
(correct to leading order) is the generalized law
\be{\xi{m}\over\eta}\approx{constant},\ee where $\eta$ decreases with
time. For orbits that spiral outward, tidal dissipation quickly
becomes negligible, so that the angular momentum parameter $\eta$
becomes nearly constant; in this case, one recovers the old result
($am\approx\xi{m}\approx{constant}$).

The analytic expressions for the critical dissipation strength
$\Gamma$, given by equations (\ref{estimate}) and (\ref{condition}),
are accurate for the physically-relevant portion of parameter space.
Nonetheless, these results must be used within their domain of
applicability. The approximations used herein break down when the
dimensionless mass loss rate becomes large ($\gamma\gg0.001$) and/or
for index combinations satisfying $q+\beta\le{p+1}$.  This treatment
only considers the dissipation of the equilibrium tide in the
convective envelope of the star; future work should also include 
tidal dissipation within the planet (although it is smaller and more
uncertain). 

\medskip 
\textbf{Acknowledgments:} We thank D. Veras for useful comments. This
work was supported by NSF Grants DMS097949 and DMS1207693.

\end{document}